\documentclass[conference]{IEEEtran}
\usepackage{cite}
\usepackage{amsmath,amssymb,amsfonts}
\usepackage{algorithmic}
\usepackage{graphicx}
\usepackage{textcomp}
\usepackage{cleveref}
\usepackage[utf8]{inputenc}
\usepackage{url}
\usepackage[singlelinecheck=false,justification=justified]{caption}
\usepackage[normalem]{ulem}

\usepackage{subcaption}
\usepackage{soul}
\usepackage{showexpl}

\def\BibTeX{{\rm B\kern-.05em{\sc i\kern-.025em b}\kern-.08em
    T\kern-.1667em\lower.7ex\hbox{E}\kern-.125emX}}

\newcommand\delete{\bgroup\markoverwith{\textcolor{red}{\rule[0.5ex]{2pt}{0.4pt}}}\ULon}

\DeclareUnicodeCharacter{0301}{\'{i}}
\begin{document}

\title{Framework for Integrating Machine Learning Methods for Path-Aware Source Routing}

\author{\IEEEauthorblockN{
Anees Al-Najjar\IEEEauthorrefmark{1}, Domingos Paraiso\IEEEauthorrefmark{2}, Mariam Kiran\IEEEauthorrefmark{1}, Cristina Dominicini\IEEEauthorrefmark{2},\\
Everson Borges\IEEEauthorrefmark{2}
, Rafael Guimaraes\IEEEauthorrefmark{2}, Magnos Martinello\IEEEauthorrefmark{3},
Harvey Newman\IEEEauthorrefmark{4}}
\\
\IEEEauthorblockA{\IEEEauthorrefmark{1}\textit{Oak Ridge National Laboratory, USA},\\
Emails: \{alnajjaram, kiranm\}@ornl.gov\\
\IEEEauthorrefmark{2}\textit{Department of Informatics, Federal Institute of Espírito Santo, Brazil}\\
Emails: domingos.paraiso@gmail.com, \{cristina.dominicini,everson,rafaelg\}@ifes.edu.br\\
\IEEEauthorrefmark{3}\textit{Department of Informatics, Federal University of Espírito Santo, Brazil}\\
Email: magnos.martinello@ufes.br\\
\IEEEauthorrefmark{4}\textit{California Institute of Technology, USA}}
Email: newman@hep.caltech.edu
}

\maketitle

\begin{abstract}

Since the advent of software-defined networking (SDN), Traffic Engineering (TE) has been highlighted as one of the key applications that can be achieved through software-controlled protocols (e.g. PCEP and MPLS). Being one of the most complex challenges in networking, TE problems involve difficult decisions such as allocating flows, either via splitting them among multiple paths or by using a reservation system, to minimize congestion. However, creating an optimized solution is cumbersome and difficult as traffic patterns vary and change with network scale, capacity, and demand. AI methods can help alleviate this by finding optimized TE solutions for the best network performance. SDN-based TE tools such as Teal, Hecate 
and more, use classification techniques or deep reinforcement learning to find optimal network TE solutions that are demonstrated in simulation. Routing control conducted via source routing tools, e.g., PolKA, can help dynamically divert network flows. In this paper, we propose a novel framework that leverages Hecate to practically demonstrate TE on a real network, collaborating with PolKA, a source routing tool. With real-time traffic statistics, Hecate uses this data to compute optimal paths that are then communicated to PolKA to allocate flows. Several contributions are made to show a practical implementation of how this framework is tested using an emulated ecosystem mimicking a real P4 testbed scenario. This work proves valuable for truly engineered self-driving networks helping translate theory to practice. 


\end{abstract}



\section{Introduction}

A truly Self-Driving Network (SDNet) requires innovative ways to merge telemetry, automation, dev-ops, and machine learning with the network infrastructure such that it can predict, change, and adapt to incoming traffic demands \cite{juniper}. While enterprises like Juniper argue that an SDNet has AI integrated into all network decisions,
little is described on how a self-driving network can practically be achieved. Networking innovations like programmable hardware, using P4 or segment routing, are bringing better network control via software \cite{KAUR2021109}. Particularly driven by the SDN vision, the path computation element (PCE) introduces a node that can compute and translate path-aware routing between source-destination pairs. Routing decisions are subject to a set of constraints, Quality of service (QoS) and policies using  MPLS (multiprotocol label switching) or segment routing to guide flows by updating the labels of the routes hops, where the traffic should transport across them. Communication between the nodes and PCE is done via PCEP (communication Protocol)\cite{ietfpcep}. However, one of the most challenging aspects of realizing the SDNet vision is little control of devices due to vendor lock-in and closed-box, where engineers can not update routing decisions easily. 

Particularly in wide area networks (WAN), these networks are witnessing massive challenges of growing complexity and varied traffic demands. Good TE routing decisions or flow models can help balance network performance, minimize congestion and optimize network utilization \cite{10.1145/3341302.3342069}, using all available bandwidth (or also referred to as run \emph{networks hotter}). However, these parameters become increasingly difficult to manage when the network size grows and becomes complex with new sources constantly being added \cite{10.1145/3563647.3563652}. In the SDN-era, TE in ISPs and Cloud WANs, have different traffic needs, delays, big transfers, etc. In science network providers, e.g., ESnet, the traffic decisions are inter-domain, scheduled and have flow requirements such as delivery on time and specific vendor protocols being used. 

In this context, traffic engineering requires path-aware networks, which allow endpoints to choose specific network paths by exposing path information at the network or transport layers \cite{rfc9217}. However, traditional table-based routing protocols often limit the ability to define specific packet network paths, leading to suboptimal routing and congestion. 
One effective method is source routing (SR) --segment routing being a prominent example-- that reduces network route states by allowing the sender to determine the packet's path through the network. This approach minimizes the need for extensive route tables and directs packets along a specified topological route. 

Using SR protocols  can help control flow allocation, but additional calculations are needed on the best paths to use. AI-enabled TE has found promising results to tackle this issue\cite{10.1145/3152434.3152441}. Techniques of machine learning to predict traffic demand matrices for better planning of traffic workloads, or learning to adapt to traffic bursts or changes in network topology can help engineers design more resilient network TE solutions without re-implementing the TE login \cite{10.1145/3563647.3563652}. Other examples include optimized network traffic management \cite{Mendiola2017, Mohammed2019, Zhang2020reinforcement, Yoo2023Deep}. These studies focus on integrating machine learning algorithms and data-driven models to predict network congestion, dynamically adjust routing paths, and optimize resource allocation.

In this paper, we present a framework for practically demonstrating path-aware routing and visualizing a self-driving network. From the AI-enabled TE side, we explore the tool Hecate \cite{hecatepaper}, which demonstrated supervised learning to help understand traffic flows and deep reinforcement learning to learn optimized network utilization results. From the path-aware network side, we explore the PolKA \cite{dominicini2020polka} architecture, which introduces a novel approach by utilizing the polynomial residue number system (RNS)  in contrast to traditional SR solutions that rely on port switching. This method enhances performance and offers advanced routing capabilities, including flexible path migration and robust failure recovery. In our proposal, various network aspects such as performance, latency and utilization are improved with APIs between Hecate and PolKA, which communicate traffic snapshots and prediction supervised learning algorithms to predict future available bandwidth on the network, returning options for PolkA to choose the best path.

Nevertheless, some challenges have to be addressed to allow an amalgamation of multiple AI capabilities (via Hecate) to be integrated with networking infrastructure (via PolKA). 
In previous works, Hecate was demonstrated in simulation, and considerable engineering is needed to translate ML decisions into practical network applications.
Secondly, PolKA previous works focused on the data plane mechanisms to enable agile path configuration, but little was done in terms of path optimization in the control plane. Finally, we must create new APIs between Hecate and PolKA, the tools to communicate traffic profiles, network topology and
current network health data to compute optimal path routing solutions, which are then dynamically allocated to the paths.


This paper proposes a novel framework that integrates PolKA source routing tool with the Hecate AI/ML optimization tool. Here, we add novel ML library APIs and new dynamic switching and routing capability between multiple flow paths to realize the first steps for autonomous optimized routing decisions. The following contributions include:
\begin{itemize}
\item Design, develop and demonstrate a smart, telemetry-driven and autonomous routing engine.
 \item Implement new supervised learning models to address autonomous routing decisions that can return path-based predictions for real-time routing decisions. e.g. Multiple regression methods.
   \item Design, develop and demonstrate source routing applications with PolKA using emulated via RARE/freeRtr testbed, allowing complex experiments to be tested before being deployed to the production systems.
\end{itemize}


The paper is organized as follows: Section~\ref{sec:background_n_motivation}  explains the importance of AI for TE and a glance about SR. The problem formulation for optimization of TE applications is presented in Section~\ref{sec:problem_formulation}. The proposed framework for optimizing SR is explained in Section~\ref{sec:Framework_Integration_for_Data-driven_Learning}. Experimental setup, implementation and evaluation of supervised ML models and PolKA experiments are presented in Section~\ref{sec:Experiment_Implementation}. Finally, comparing the proposed contribution with state of the art, and concluding our findings with potential future directions are demonstrated in Section~\ref{sec:related_work} and Section~\ref{sec:conclusion_n_future_work}, respectively.
 
\section{Background and Motivation}
\label{sec:background_n_motivation}

\subsection{Need for Traffic Engineering and AI}

As network traffic grows at exponential rates \cite{10.1145/3503954.3503958}, providers often cap network bandwidth at 40\% to reduce hotspots or traffic bursts, to prevent congestion and packet loss \cite{b4}. However, this leads to a highly underutilized network and continually upgrading the network with more capacity is an expensive solution. As AI helps solve complex problems such as self-driving cars, and complex protein structures, it is also been explored to help with traffic
engineering \cite{hecatepaper}. Studying traffic patterns from network monitoring data, TE solutions estimate traffic profiles and traffic demands to come up with innovative solutions for flow allocation. Previous
works DeepRoute \cite{10.1007/978-3-030-45778-5_20} and \cite{hecatepaper} use an AI agent using greedy
Q-learning to learn optimal routing
strategies to minimize flow completion time. However, there are needs to be an automated way in which these decisions can be communicated on the network. Hecate leveraged data-driven learning to compute path optimization using two approaches (1) optimizing routing configurations by predicting
future traffic conditions depending on past traffic
patterns or (2) optimizing routing configurations based on the
number of feasible traffic scenarios to improve performance
parameters \cite{10.1145/3152434.3152441}. Using near real-time data on bandwidth, jitter, or latency, one can predict future network performance and traffic demand matrix to infer the best paths between source-destination pairs. However, this approach has limitations in dynamic large network topology as (1) networks grow from 10s to 100s of routers, and (2) the traffic demand matrix changes. It is important to provide an end-to-end solution that can translate the AI decisions directly into the network to help improve the network capacity challenges. The proposed integration framework, presented in Sec.~\ref{sec:Framework_Integration_for_Data-driven_Learning}, leverages AI solutions to optimize traffic allocation and SR using PolKA based on end-to-end network stats monitoring.

\subsection{Source Routing}

The benefits of SR over traditional table-based routing include a reduction in network states and the optimal use of network capacity \cite{sourcey}. The most common method of implementing SR is Port Switching, where the route label represents an ordered list of output ports (or network addresses). Each hop executes the forwarding operation by popping the first element of the list \cite{sourcey}, necessitating an update to the route label in the packet at each hop.

On the other hand, PolKA SR utilizes the residue number system (RNS) with Galois field (GF) polynomials of order 2 to derive the route label, which remains unchanged throughout the entire path \cite{dominicini2020polka}. In this scheme, at any core node, the output port (portID) is determined by the remainder of the binary polynomial division (i.e., a mod operation) of the route identifier (routeID) of the packet by the node identifier (nodeID). This polynomial mod operation is enabled in programmable switches by reusing the cyclic redundancy check (CRC) hardware \cite{dominicini2020polka}.

\begin{figure}[!t]
        \centering
        \includegraphics[width=0.48\textwidth]{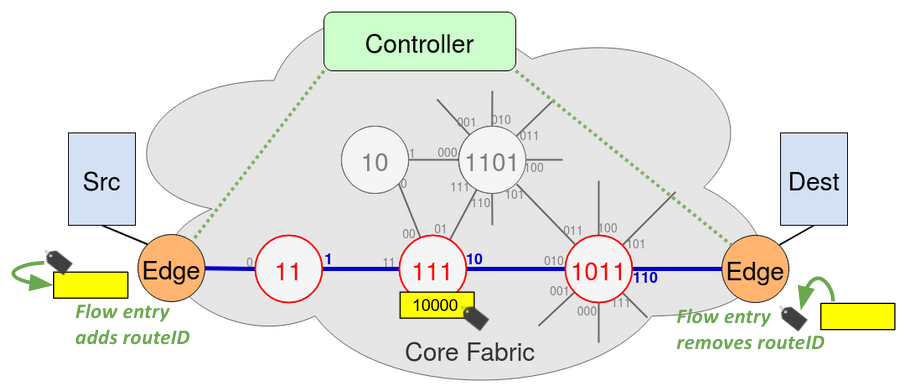}
        \captionof{figure}{PolKA source routing tool uses polynomial identifier to guide the packet through the network.}
        \label{fig:polka_example}   
\end{figure}

For example, consider a path through three core nodes as shown in Fig.~\ref{fig:polka_example}, each with its own polynomial identifier: $s_1(t) = t + 1$, $s_2(t) = t^2 + t + 1$, and $s_3(t) = t^3 + t + 1$. The output port polynomials for these nodes are $o_1(t) = 1$, $o_2(t) = t$, and $o_3(t) = t^2 + t$. The route identifier (routeID) for this path is computed using polynomial Chinese Remainder Theorem (CRT) and is embedded in the packet by the controller\cite{borges2024pot}. As the packet traverses the nodes, each node computes its output port by dividing the routeID by its polynomial identifier. For instance, if routeID is $10000$, node $s_2(t)$ calculates the port label by finding the remainder of $10000$ divided by $t^2 + t + 1$, resulting in port label $2$. By following the polynomial identifier, the packet can be guided through the network, taking paths as the packet arrives at the next hop.

\section{Problem Formulation}
\label{sec:problem_formulation}

Traffic engineering involves developing models based on traffic profiles or demands. These have to be re-engineered when any changes in topology, traffic patterns, or users have happened. Flow management approaches use flow tables, as in OpenFlow switches, that calculate the forwarding paths. These models help the tradeoff between latency and load balancing.

TE has been extended to routing protocols such as MPLS to achieve a few of the desired properties such as (1) maximum usable bandwidth or reservation allowed, (2) unreserved bandwidth available to use, or (3) traffic engineering metric for special flow characteristics. To understand how optimal paths can be computed, we need to describe the network flow model problem. This is a combinatorial optimization problem, where the network computes edges and paths based on the capacity constraints, and incoming and outgoing flow. These include challenges of maximum flow problems, minimum-cost flow problems or multiple optimization parameter problems. In ISPs, we see the min-max problem as the main flow computation problem, which essentially helps maximize flows while minimizing congestion.

\subsection{Optimizing min-max problem}

\begin{figure}
  \centering
    \includegraphics[width=0.8\linewidth]{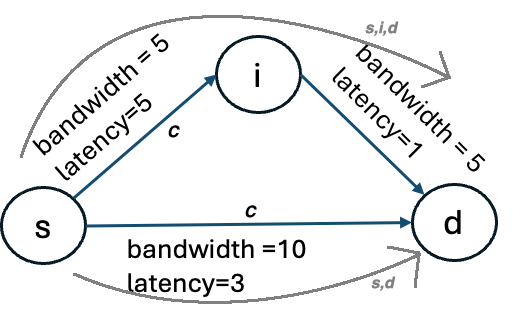}  \caption{Simple network setup with multiple routes having different QoS parameters.}
  \label{fig:simple}
\end{figure}

As flows are allocated in a network, their optimal allocation can allow better utilization of resources. For example, Fig.~\ref{fig:simple} shows a simple 3-node network where links have bandwidth and latency specifics. Arriving flows at source `s' are traveling to destination `d' and need to be allocated to minimize congestion or maximize a link utilization at a given time. We represent the flows as the demand volume between $(s,d)$. 

Since the demand volume can be divided among two paths, we can write:
\begin{equation}
    x_{sd}+x_{sid}=h
\end{equation}
while $0<=x_{sd}<c$ and $0<=x_{sid}<c$. With latency being an additional factor to optimize, the system can have infinite possible solutions that satisfy the latency, bandwidth, and capacity constraints. Thus the total cost of allocating demand flows becomes our objective function for routing:
\begin{equation}
    minimize_{\{x_{sd},x_{sid}\}} F=\xi_{sd}x_{sd}+\xi_{sid}x_{sid}
\end{equation}

However, we also have load balancing, variation in capacity, and delay to be considered. 
The link utilization is defined as the amount of flow on a link divided by the capacity of that link, written as $\frac{x_{sd}}{c}$, then the maximum utilization is written as $max(\frac{x_{sd}}{c},\frac{x_{sid}}{c})$.

With a delay, the objective function finds the optimal capacity to minimize the delay. Using the arrival rate for demand flow, we can rewrite the delay as follows:
\begin{equation}
    minimize_{x} F=\frac{x_{sd}}{c-x_{sd}}+\frac{2x_{sid}}{c-x_{sid}}
\end{equation}
Using the above conditions, the problem of finding an optimal objective function becomes a Linear Programming (LP) problem, with all constraints being linear functions. This can be solved using LP solvers but can become increasingly complex with more variables to consider. These foundations have been adapted from \cite{deepbook}.

In addition to finding an optimal objective function, we have dynamic variables such as network QoS parameters, jitter, loss, and varying occupied bandwidth. Most flows come with specific requirements with guaranteed bandwidth or QoS parameters. Developing an ML solution can help learn optimal objective functions from the data being collected in real-time.



\emph{Limitations in Flow Tables:} Flow-based routing or aggregate routing can also help in grouping flows to follow specific rules. However, these are still limited by table size and how to manage multiple flow tables. Replacing the flow tables with SR and AI/ML methods, where more information can be held can help optimize the flow table limitations.

\emph{Real-time Decision Making:} Leveraging the current QoS values (at $t_i$) of the routes makes us aware of only the current routes' status with no insight into the QoS patterns of the topology routes. Allocating the network traffic based on the current QoS status of the route may affect the allocated flows due to unexpected network impairment factors, and as a result, the probability of dropping the flows is higher. Hence, it is important to utilize the history of topology routes to estimate the QoS parameter of routes for $t_{i+x}$, where $1<x<n$. For this reason, we utilize AI/ML models for path-aware QoS of the topology routes to support PolKA for optimal routing decisions.

We leverage ML regression as part of Hecate integrated framework presented in Fig.~\ref{fig:POLKA-Hecate_Framework}. The regression model will extrapolate the QoS parameters for the topology routes based on the previous measurements. Several regression models are explored to choose the best performance model to be integrated into the proposed framework.

\section{Framework Integration for Data-driven Learning}
\label{sec:Framework_Integration_for_Data-driven_Learning}

Using SR, when a flow or packet enters the network, it can be assigned a route label that guides each router along the path, specifying which interface to use for processing. This approach lessens the burden on a centralized controller and enables decentralized routing decisions as packets traverse the network. It offers several advantages, including network simplification, increased resilience, and improved scalability. However, this method also introduces challenges related to traffic engineering and optimization. In this discussion, we address these issues and explore potential solutions.

\begin{figure}[t]
        \centering
        \includegraphics[width=0.5\textwidth]{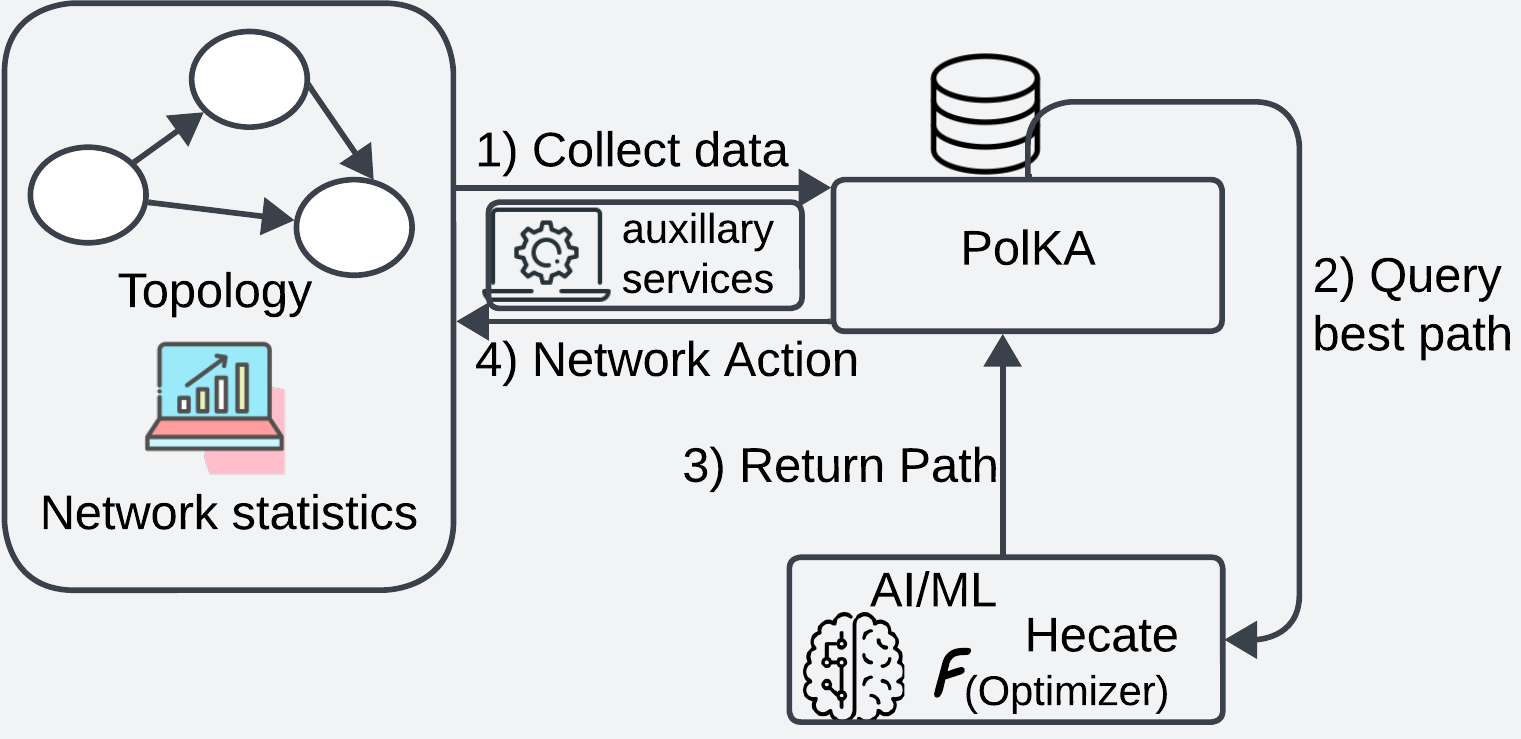}
        \captionof{figure}{PolKA-Hecate integration framework}
        \label{fig:POLKA-Hecate_Framework}   
\end{figure}

Fig.~\ref{fig:POLKA-Hecate_Framework} shows how Hecate and PolKA framework exchange information for a data-driven path routing problem. The framework --via leveraging PolKA service-- is responsible for managing the network and ensuring efficient data flow. This interaction between PolKA and Hecate enables \textit{adaptive routing}. The diagram also depicts the interaction among auxiliary framework services, including the Controller, Scheduler, and Telemetry Services, highlighting their roles in managing network flows. Hecate adopts the best-performing regression model, which is then integrated into the routing framework shown in Fig.~\ref{fig:POLKA-Hecate_Framework}. The ML model predicts QoS at time $t_{i+1}$. Multiple regressors are explored in Section~\ref{sec:Experiment_Implementation} and the best performance model is integrated into the routing framework. The path QoS estimations are sent to the Optimizer, which selects the optimal route based on the defined objective function. 

\begin{figure*}[t]
        \centering
        \includegraphics[width=0.8\textwidth]{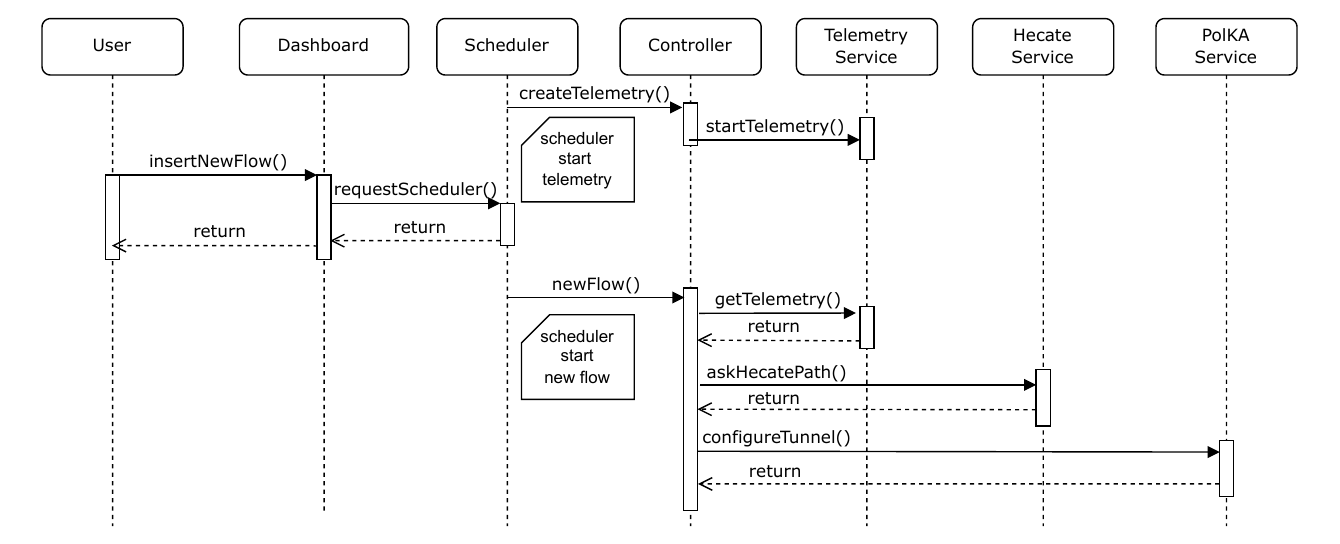}
    \caption{\centering Sequence diagram of Hecate-PolKA framework }
  \label{fig:sequencediagram}
\end{figure*}

The sequence diagram (Fig.~\ref{fig:sequencediagram}) illustrates the operations of managing interactions across various framework components. To maintain continuous monitoring and management of network traffic, the system offers visual feedback through link occupation graphs displayed on the Dashboard. This allows for ongoing observation of network performance and the efficiency of path allocations. When a user requests a new flow via the Dashboard, the request is sent to the Scheduler. 

The path allocation process for each new flow starts when the Scheduler notifies the Controller of the intent to establish a new connection. To determine the optimal path, the system first gathers telemetry data from the network. At predefined intervals, the Controller activates agents to collect telemetry data from relevant network paths, focusing on metrics like flow rate and latency as outlined in the topology description. This data is then transmitted to the Telemetry Service, where it is stored in a time series database for analysis.

The Controller retrieves the stored telemetry data, formatted as a dataset of time-indexed values, and provides it to the Optimizer. When a new data flow arrives, the Controller consults the Optimizer to determine the most suitable path. After the optimal path is identified, the Controller communicates this decision to the SR Service, establishing the path and configuring a policy to route the flow through it by adjusting the edge routers. In this setup, Hecate is the Optimizer, while PolKA is the SR routing mechanism.



\section{Experiment Implementation}
\label{sec:Experiment_Implementation}
\subsection{Training Hecate: Data-driven Learning}

\subsubsection{Dataset} 
\begin{figure}[t]
\centering
\begin{subfigure}[c]{0.6\textwidth}
   \includegraphics[width=0.7\textwidth,height=1.5in]{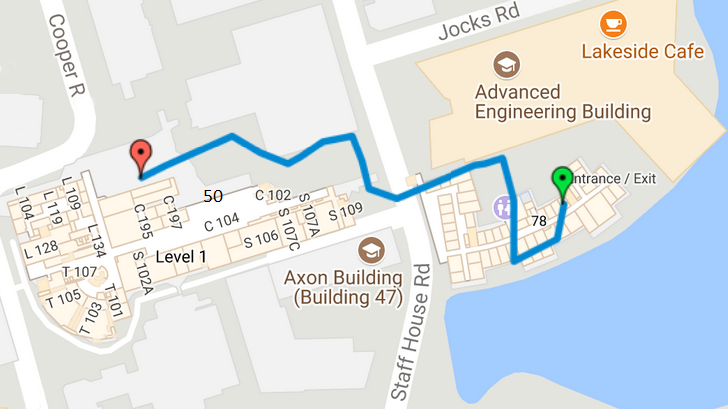}
\caption{{Selected path for measuring wireless bandwidth at UQ}}
\label{subfig:Selected_Path_for_Measuring_Wireless_Bandwidth_UQ}
\end{subfigure}
\hfill
\begin{subfigure}[c]{0.6\textwidth}
        \includegraphics[width=0.7\textwidth,height=1.5in]{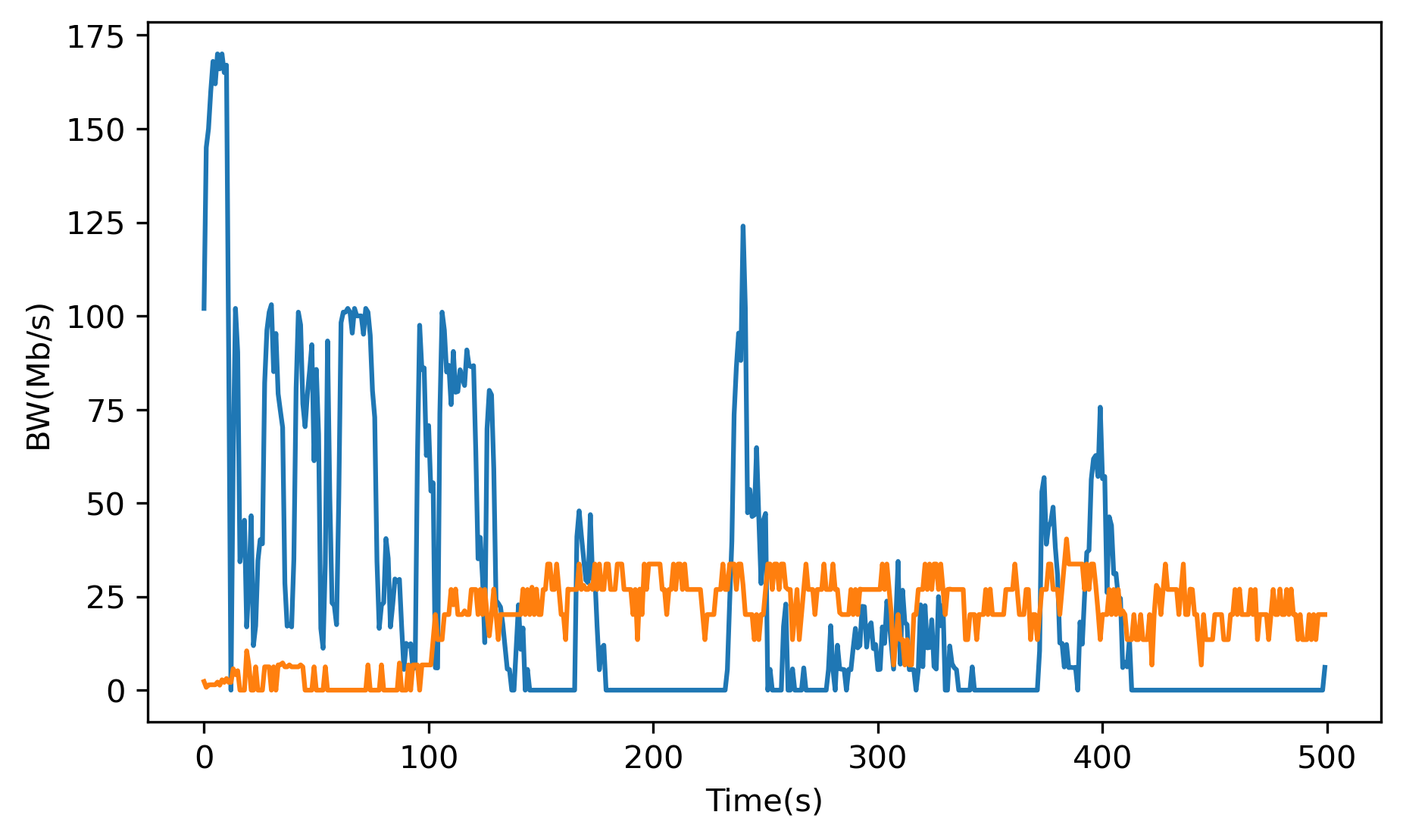}
\caption {{WiFi (Path 1) vs LTE (Path 2) measured bandwidth}}
\label{subfig:WiFi_vs_LTE_Measured_Bandwidth}
\end{subfigure}
\caption{{Wireless bandwidth measurement of LTE and WiFi over a selected path at UQ}}
\label{fig:Uq_dataset}
\end{figure}

For proof of concept, we utilize a real dataset to test and validate the ML models integrated with Polka to optimize route selection. We use a Wireless dataset previously collected over a dedicated path at The University of Queensland (UQ) in June 2017 \cite{al2020network,al2018enhancing}. This dataset includes bandwidth measurements of LTE and WiFi measured using Iperf~\cite{iperf3} installed on two laptops and collected on a second-unit basis for 500 seconds. The experiment began indoors at building 78, and the experimenter moved outdoors to complete the experiment at building 50, as depicted in Fig.~\ref{subfig:Selected_Path_for_Measuring_Wireless_Bandwidth_UQ}. The measured bandwidth demonstrates variation based on the network wireless type and the experiment location, whether it was conducted inside or outside the, as shown in Fig.~\ref{subfig:WiFi_vs_LTE_Measured_Bandwidth}. For instance, The WiFi channel supports better bandwidth if the experiment is conducted indoors (from time 0 to 100); on the contrary, the LTE wireless network measured very low bandwidth during the same time. We use this variation in bandwidth across two paths of a selected topology with two path routes to choose from.

\subsubsection{Supervised ML and Prediction Methods}
\label{subsec:Supervised_ML_and_Prediction_Methods}

We investigated eighteen ML Regressors (R) to estimate the bandwidth of a network topology and feed the routing framework for optimal routing decisions. The regressors are alphabetically sorted as follow: Ada Boost Regressor (R1:AdaBoostR), ARD Regression (R2:ADAR), Bagging Regressor (R3:Bagging), Decision Tree Regressor (R4:DTR), Elastic Net (R5:ElasticNet), Gradient Boosting Regressor (R6:GBR), Gaussian Process Regressor (R7:GPR), Histogram-based Gradient Boosting Regression (R8:HGBR), Huber Regressor (R9:HuberR), Lasso (R10:Lasso), Linear Regression (R11:LR), RANdom SAmple Consensus Regressor (R12:RANSACR), Random Forest Regressor (R13:RFR), Ridge (R14:Ridge), Stochastic Gradient Descent Regressor (R15:SGDR), Support Vector Machine/Linear Kernel (R16:SVM-Linear), Support Vector Machine/RBF Kernel (R17:SVM-RBF), and Theil-Sen Regressor (R18:TheilSenR). Based on their performance, the best regression model will be integrated with the routing framework (Fig.~\ref{fig:POLKA-Hecate_Framework}) to predict the end-to-end bandwidth.

\subsection{ML Experimental Evaluation}
\label{subsec:ML_experiment_setup}
The ML modules used with the proposed routing framework (Fig.~\ref{fig:POLKA-Hecate_Framework}) are developed under Python v3.10 programming language. We leveraged scikit-learn package to instantiate and run the regression models described above. These models are executed with the default hyperparameters. In addition, we use the Mean Squared Error (RMSE) as a performance evaluation metric of the regression models mentioned earlier.

Several steps are taken to configure a certain ML pipeline for QoS estimation. Initially, we proportionally split UQ dataset into training and testing sets by 75\% and 25\%, respectively. Then, the dataset will be transformed into a normal distribution to avoid improper results from the applied regression models. 
Regressors may expect normal distributions of the dataset features, in this case, LTE (Path 1) and WiFi (Path 2) of the UQ dataset. Hence, we used the StandardScaler utility function to re-scale the dataset features, where it calculates the mean and standard deviation of the dataset features at the training set, using \textit{fit} method, and then scales the testing set using \textit{transform} method. As a later operation after the ML model is applied, \textit{inverse\_transform} on the estimated values are applied to get the feature values back to their original scale.

After that, we applied the ML models to estimate the bandwidth of different paths. We set the history of measurements used in the regression models to 10 values that represent $t_i\ to\ t_{i-9}$. These values are passed to the models to predict bandwidth at $t_{i+1}$. The training dataset is further split to fit the models based on the historical values, while the testing dataset is utilized for predicting $t_{i+1}$ values

\begin{figure}
  \centering
    \includegraphics[width=\linewidth]{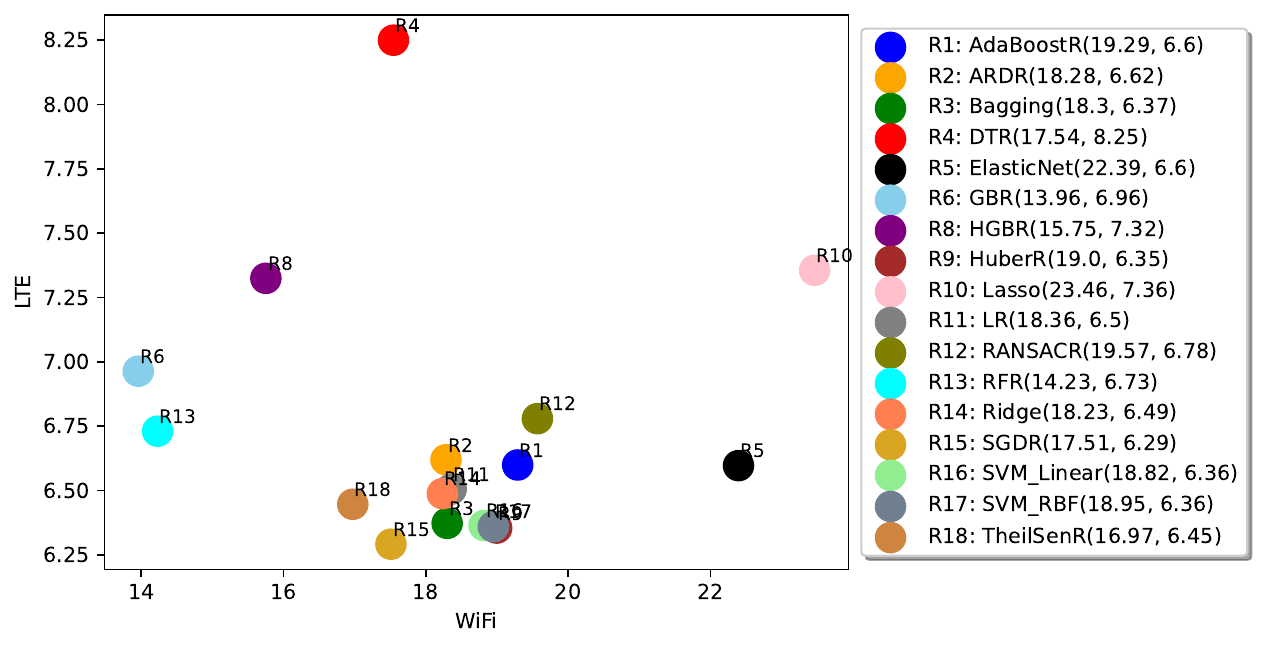}  \caption{RMSE of multiple regression models applied on the bandwidth of Path 1 and 2. It shows the models towards zero on the X and Y axes have  better performance.}
  \label{fig:RMSE_Regressors}
\end{figure}

\begin{figure}[t]
\centering
\begin{subfigure}[c]{0.6\textwidth}
   \includegraphics[width=0.7\textwidth,height=1.5in]{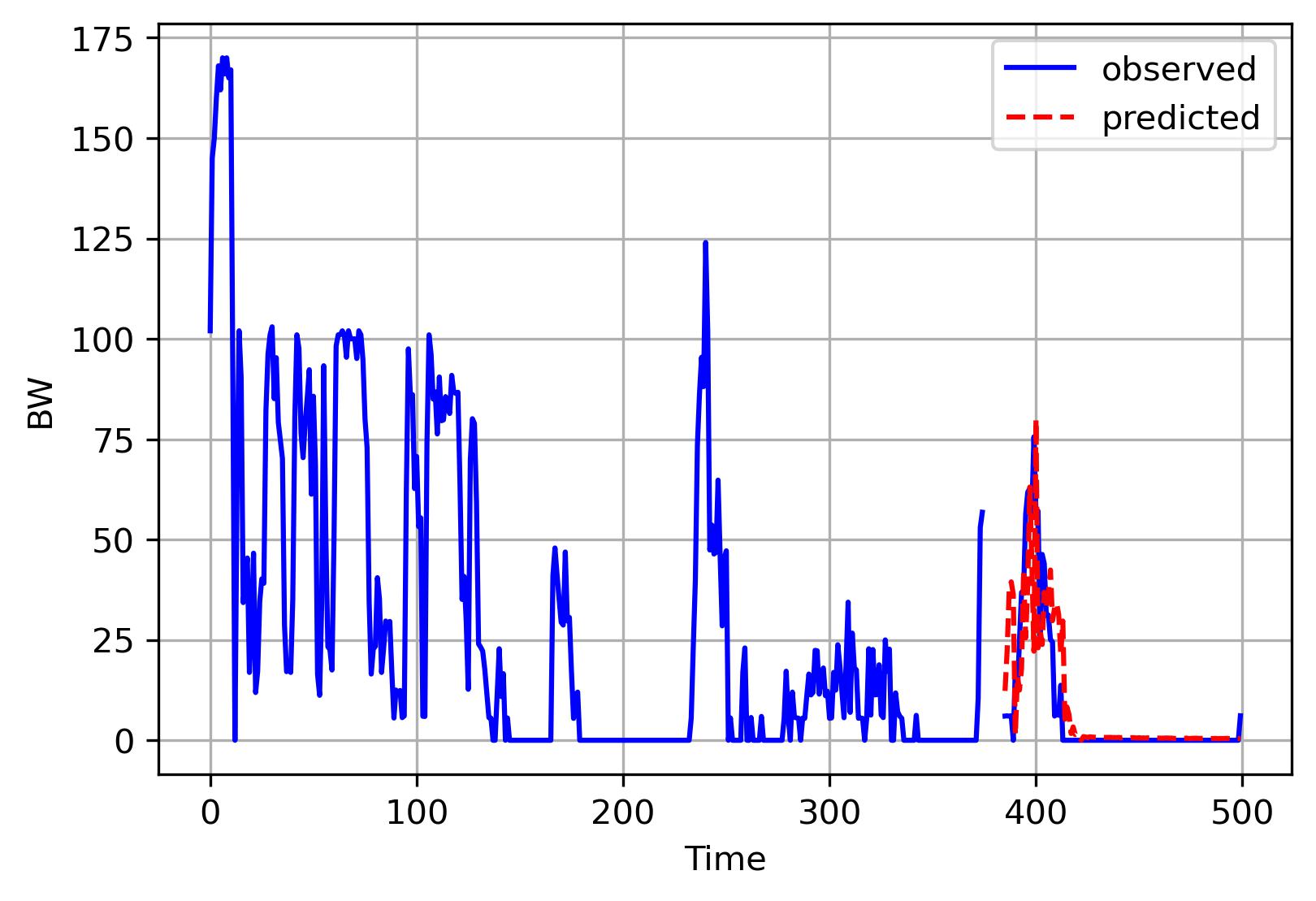}
\caption{{Observed and predicted WiFi (Path 1) bandwidth}}
\label{subfig:RandomForestRegressor_WiFi}
\end{subfigure}
\hfill
\begin{subfigure}[c]{0.6\textwidth}
        \includegraphics[width=0.7\textwidth,height=1.5in]{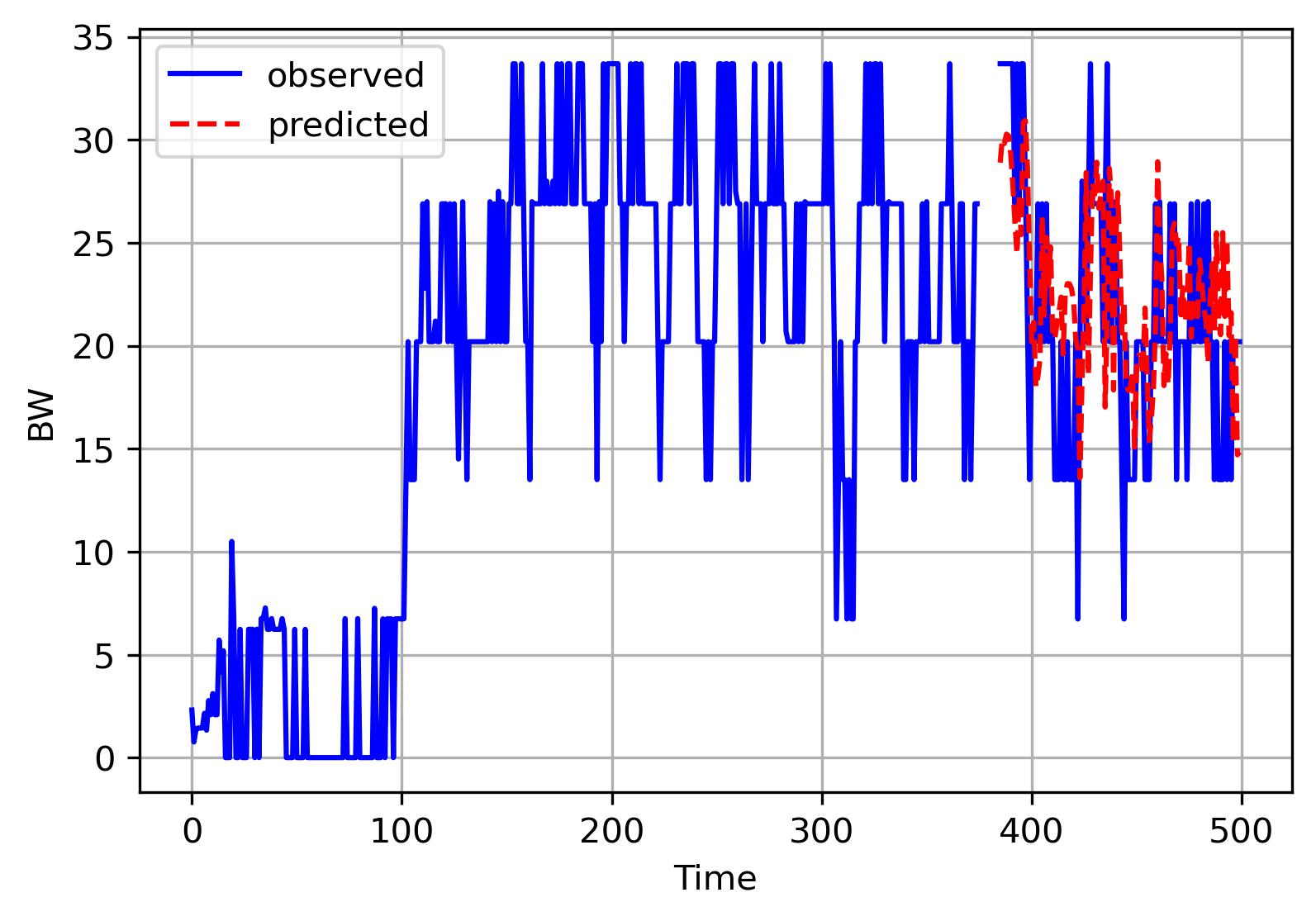}
\caption {{Observed and predicted LTE (Path 2) bandwidth}}
\label{subfig:RandomForestRegressor_LTE}
\end{subfigure}
\caption{{Random Forest Regression model to predict routes bandwidth}}
\label{fig:RandomForestRegressor}
\end{figure}

\begin{figure}[t]
\centering
\begin{subfigure}[c]{0.6\textwidth}
   \includegraphics[width=0.7\textwidth,height=1.5in]{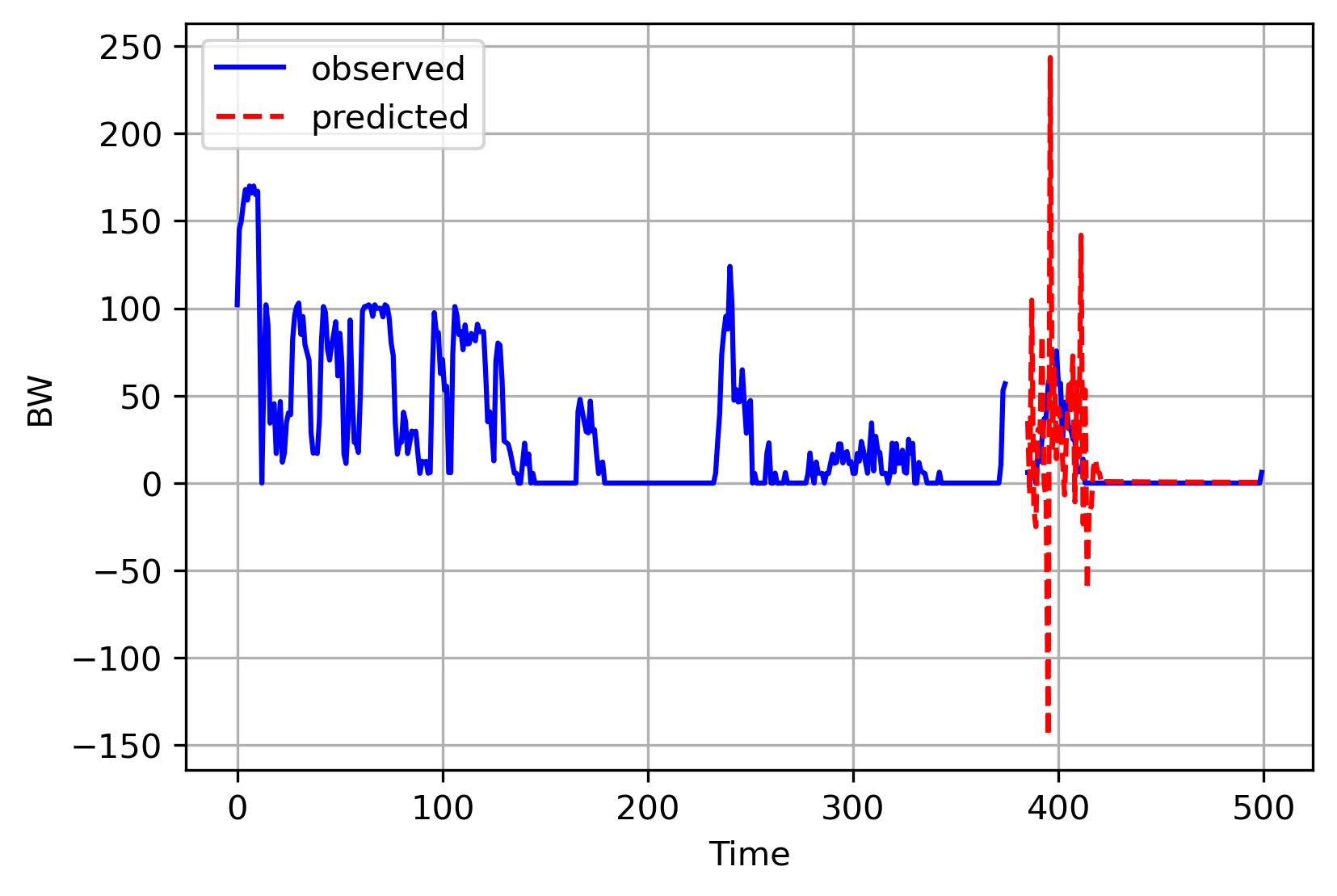}
\caption{{Observed and predicted WiFi (Path 1) bandwidth}}
\label{subfig:GaussianProcessRegressor_WiFi}
\end{subfigure}
\hfill
\begin{subfigure}[c]{0.6\textwidth}
        \includegraphics[width=0.7\textwidth,height=1.5in]{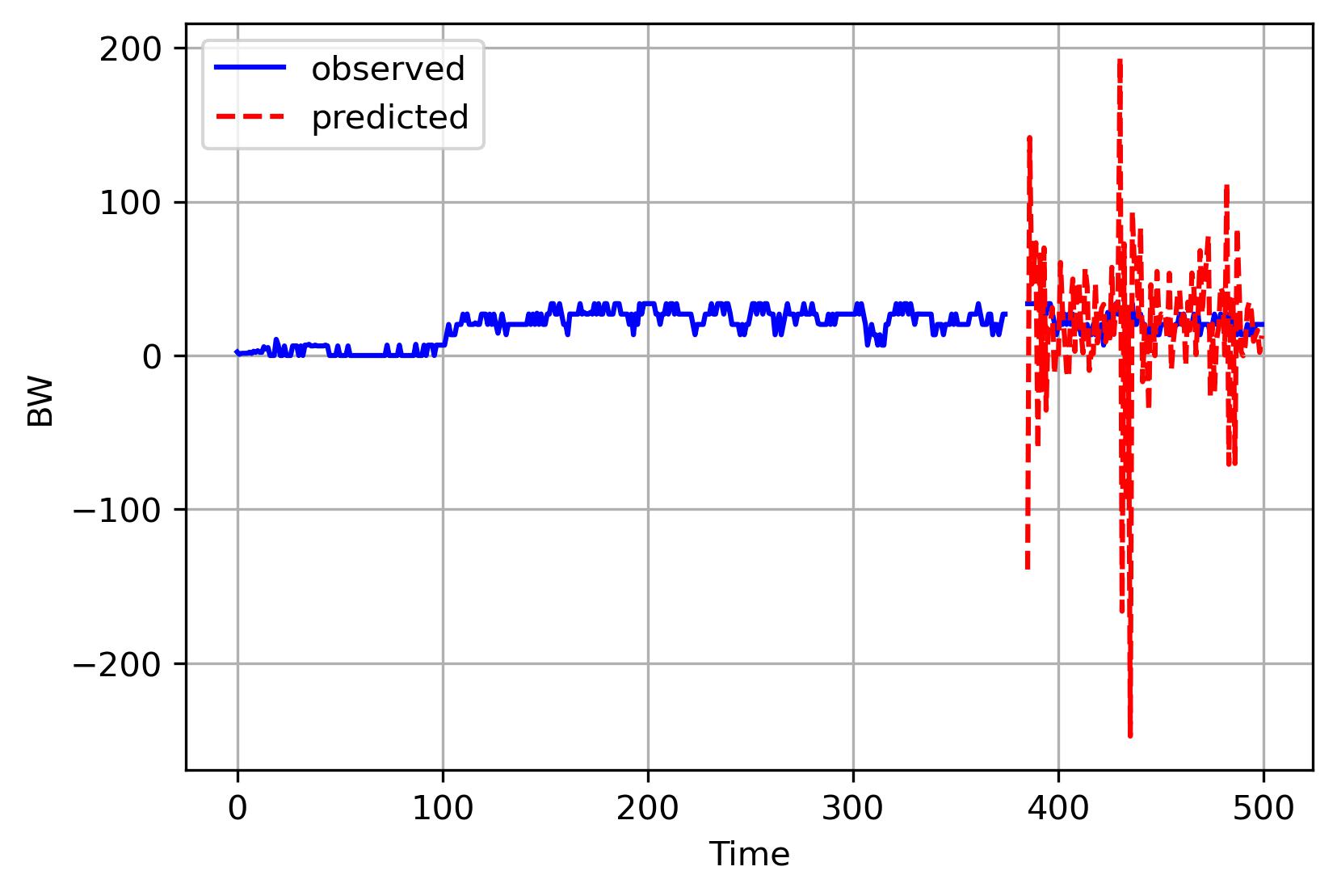}
\caption {{Observed and predicted LTE (Path 2) bandwidth}}
\label{subfig:GaussianProcessRegressor_LTE}
\end{subfigure}
\caption{{Gaussian Process Regression model to predict routes bandwidth}}
\label{fig:GaussianProcessRegressor}
\end{figure}

The performance of the regression models is presented as a scatter plot, as shown in Fig.~\ref{fig:RMSE_Regressors}, that depicts the RMSE values of WiFi (Path 1) and LTE (Path 2) as Cartesian coordinates, where X-axis and Y-axis represent the RMSE for WiFi and LTE, respectively. GPR is excluded from the scatter plot due to the high RMSE values belonging to WiFi (34.75) and LTE (52.43). As can be seen from the figure, RFR and GBR are the best regression models with the lowest RMSE, whereas other models demonstrate higher RMSE for different paths, especially when they get further from the lower left corner of the plot. Based on the plot representation, we chose RFR (WiFi: 14.23, LTE: 6.73) as the regression model to be integrated with the routing framework.

After figuring out the performance of these ML models, we validate the predicted versus the observed bandwidth of different paths of the best model (RFR) and the worst model (GPR), as shown in Figs.~\ref{fig:RandomForestRegressor}, and~\ref{fig:GaussianProcessRegressor}, respectively. We noticed that the RFR predicts bandwidth for WiFi~(Fig.~\ref{subfig:RandomForestRegressor_WiFi}) and LTE (Fig.~\ref{subfig:RandomForestRegressor_LTE}) very close to the observed real bandwidth. On the contrary, we notice a big variation between the observed and predicted bandwidth of WiFi(Fig.~\ref{subfig:GaussianProcessRegressor_WiFi}) and LTE(Fig.~\ref{subfig:GaussianProcessRegressor_LTE}) when GPR model is used.

Hecate computes the predicted values for the next 10 steps and returns the best path, where the most available bandwidth is as a recommendation for PolKA to use. This way we ensure the flows will have less congestion points in the future when they are allocated to these paths.

\subsection{PolKA enabling path-aware networks}
\label{subsec:Utilizing_Polka_Emulation_and_freeRtr}


The PolKA routing approach has been tested on commercial programmable equipment in international testbeds for high-speed data transfers (10 Gbps and 100 Gbps), showing similar performance to traditional forwarding methods \cite{dominicini2021deploying, borgessc22}. It was later integrated in an emulated testbed with the RARE/freeRtr platform \cite{wtestbeds2022} for automated control plane operations and simplified tunnel creation \cite{borges2022lifecycle}. 

We extended the virtual environment proposed in \cite{wpeif}, using the RARE/freeRtr platform to prototype and integrate new protocol features before deployment on a large-scale physical testbed with programmable P4 switches. In this section, we showcase two experiments on how PolKA can agilely configure the paths informed by an optimization engine to perform resource allocation decisions in path-aware networks. We describe the virtual environment, the topology scenario, the experiments and the results. The same principles can be used in future works when the integration between PolKA and Hecate is fully completed.

\subsubsection{Virtual testbed with freeRtr}

RARE/freeRtr is a network routing software that supports both traditional and new protocols, such as PolKA \cite{wtestbeds2022}. It can be used to emulate networks or as a control plane for hardware devices and supports various data planes, including DPDK and P4. 

The testing environment used was a computer with an Intel i7 processor, 12GB of RAM, running Linux Debian 11, and VirtualBox 7. 
We created a template VM in VirtualBox with the following installation: Debian 11, RARE/freeRtr, iperf3, and bwm-ng. This VM was subsequently cloned, and scripts were used to automate the customization of RARE/freeRtr configuration files for each node. 

\begin{figure}[!t]
        \centering
        \includegraphics[width=0.48\textwidth]{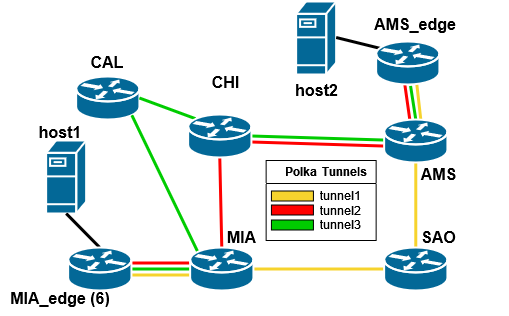}
        \captionof{figure}{P4 testbed topology in emulated environment}
        \label{fig:topology}   
\end{figure}

The topology used in this work represents a subset of the nodes from the Global P4 Lab testbed\footnote{https://wiki.geant.org/pages/viewpage.action?pageId=609058868}, which has P4 programmable switches in Europe, the United States, and Brazil, as shown in Fig.~\ref{fig:topology}. To emulate this topology, we created 9 Virtual Machines (VMs) with 1GB of RAM running Debian 11. On the VMs functioning as routers, we installed RARE/freeRtr. We configured the interfaces between the MIA and SAO routers to use physical machine Ethernet interfaces and introduced a 20ms delay in the host operating system using the \textit{tc} command. To limit the transmission rate, we used a native VirtualBox feature that allows setting different limits on network interfaces.
To emulate the topology, we used the internal network feature provided by VirtualBox. 

\begin{figure*}[t]
  \centering
    \includegraphics[width=\textwidth]{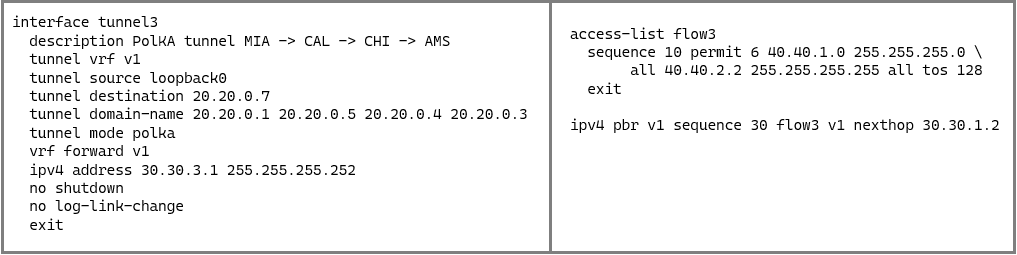}
    \caption{Configuration setup for PolKA on freeRtr platform}
  \label{fig:Configuration}
\end{figure*}

PolKA tunnels, access control, and Policy-Based Routing (PBR) are configured on the edge routers according to the freeRtr commands shown in Fig.~\ref{fig:Configuration}. In the example, the \textit{access-list} section specifies that network 40.40.1.0/24 can access machine 40.40.2.2 using protocol 6 (TCP). The ToS (Type of Service) indicated at the end of the command filters only packets with that indication. The tunnel is created in the `interface` section, and the configuration \textit{tunnel destination 20.20.0.7} specifies that the tunnel goes to the AMS edge router, while \textit{tunnel domain-name} provides the list of routers that are part of the explicit path, which will be internally converted by freeRtr into a PolKA routeID to be encapsulated in the packets passing through the tunnel. In the last line, a PBR is created indicating that access control \textit{flow3} will use tunnel 3, as the address 30.30.3.2, which is the IP on the other side, is specified.

The framework uses a message queue system to facilitate communication between its components described in Section~\ref{sec:Framework_Integration_for_Data-driven_Learning}. It includes functions for managing edge routers, allowing it to control PolKA tunnels, access lists, PBR, and telemetry data. When a new flow is introduced, the framework queries Hecate to determine the optimal path, and then directs the flow accordingly, as described above. Additionally, the framework periodically collects telemetry data from the tunnels, which is stored for future queries to Hecate. 
Using this framework, we manage FreeRtr configurations by sending messages through a Message Queue to reconfigure the router. A service receives these messages, applies the necessary commands to reconfigure FreeRtr, and then ensures the router operates with the updated configuration, as shown in Fig.~\ref{fig:Configuration}.


\subsubsection{Experiments}



\begin{figure}[!t]
        \centering
        \includegraphics[width=0.48\textwidth]        {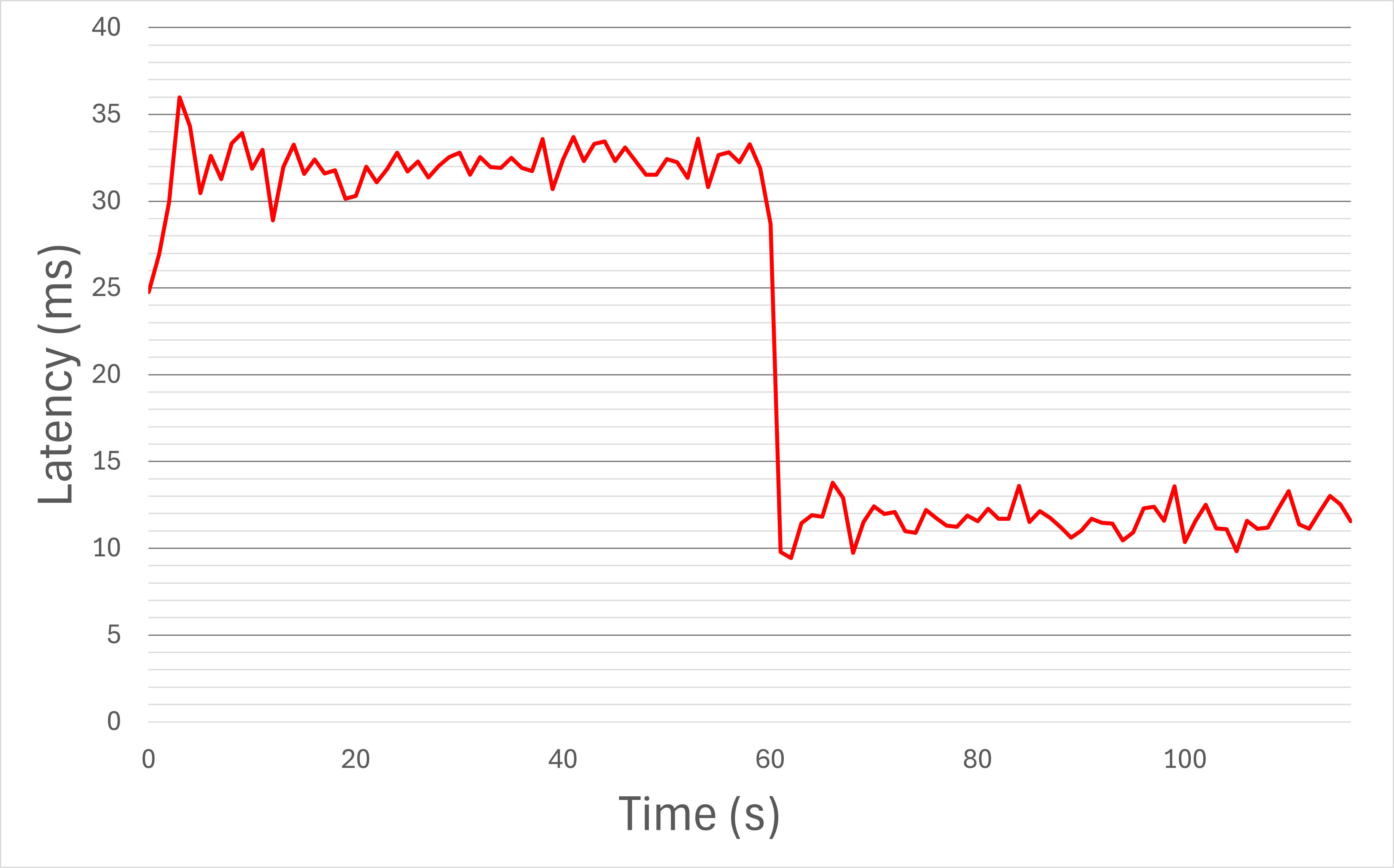}
        \captionof{figure}{Agile migration to a path with lower latency.}
        \label{fig:test1result}
\end{figure}

\begin{figure}[!t]
        \centering
        \includegraphics[width=0.48\textwidth]        {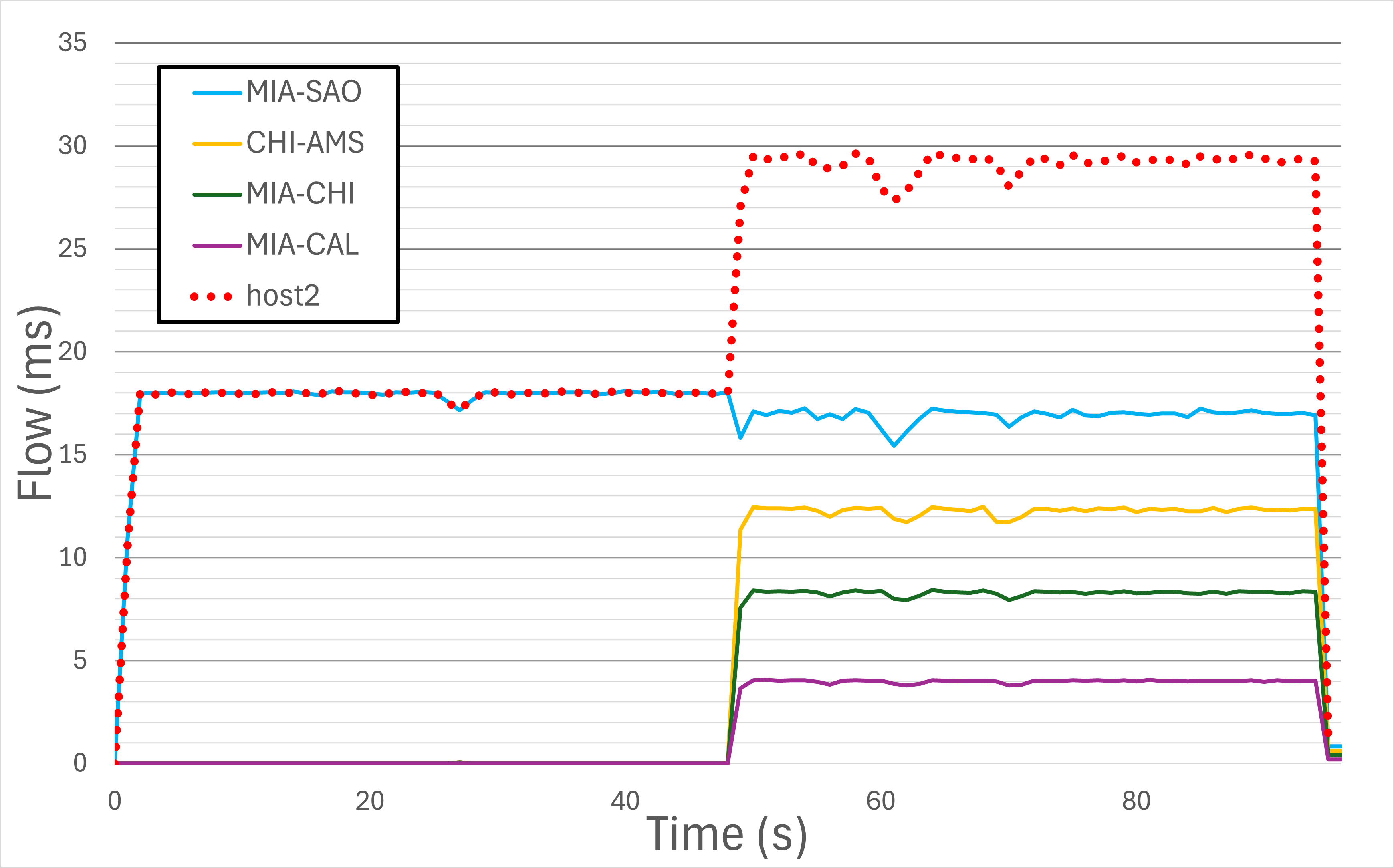}
        \captionof{figure}{Flow aggregation with multiple paths to increase available bandwidth.}
        \label{fig:test3result}   
\end{figure}

We consider two phases: (i) the controller allocates the flow to an arbitrary path; (ii) the controller consults an optimization engine that is able to improve the previous allocation decision. We want to demonstrate that, once the path is selected, PolKA allows the agile selection of a configured tunnel at the network edge through PBR, without any modifications to the core network.
To this end, we consider two experiment scenarios:
\begin{itemize}
    \item The first experiment (Fig.~\ref{fig:test1result}) demonstrates the agile migration to a lower latency path to optimize the performance of a specific flow. We configured the PBR to direct the flow through the tunnel (MIA-SAO-AMS) for 1 minute, during which the \textit{ping} command sent ICMP packets between host1 and host2. Then, we request a path allocation solution for latency minimization to the optimizer, which returns path MIA-CHI-AMS. Finally, we changed the PBR to route the flow through this new tunnel (MIA-CHI-AMS), improving the user perceived latency.
    \item The second experiment (Fig.~\ref{fig:test3result}) demonstrates how distributing flows across different paths can avoid network limitations. To emulate varying capacities, we restricted the bandwidths of the links: MIA-SAO, SAO-AMS, and CHI-AMS to 20 Mbps, MIA-CHI to 10 Mbps, and MIA-CAL and CAL-CHI to 5 Mbps. We generated three TCP flows with different Types of Service (ToS) between host1 and host2. Initially, all flows used Tunnel 1, resulting in a maximum throughput of less than 20 Mbps. Then, we request a path allocation solution, considering a bandwidth metric. The result is the modification of one flow to Tunnel 2 and another to Tunnel 3. Fig.~\ref{fig:test3result} shows the average throughput and indicates an increase in total throughput (30 Mbps) as the flows traverse different paths to reach the final host. This mechanism can be used for both flow aggregation and load balancing.    
\end{itemize}

It is important to note that, in these two scenarios, each path migration is triggered by a single modification of a PBR entry in the ingress edge node (MIA\_edge). 


\section{Related Work}
\label{sec:related_work}
Networks are desired to have the ability to actively
steer traffic flows across predetermined paths to
efficiently utilize resources. This is normally done through protocols such as IP routing protocols, MPLS
or RSVP-TE \cite{hecatepaper}. While these provide some solutions, these cannot be scaled to larger network topologies and cost engineering time. MPLS-TE or Segment Routing can make it easy to move traffic and integrate it into existing solutions. 

Various versions of TE optimizers have tackled Cloud and ISP WAN, for both inter and intra-domain optimization. However, as objectives continue to evolve, multiple designs can be tailored for the WAN being worked with \cite{10.1145/3563647.3563652}. 
A few examples include Google's B4 \cite{b4}, SWAN, Teal \cite{10.1145/3603269.3604857}, and RADWAN \cite{10.1145/3230543.3230570}. These approaches use a traffic demand matrix to calculate optimal bandwidth allocation, sometimes using machine learning or rate-adaptive techniques to dynamically change the bandwidth limits to achieve better traffic throughput.

Researchers have also optimized TE challenges in the application-layer traffic optimization (ALTO) protocol, providing information about application performance and resource utilization as interfaces between client and server. The server provides cost maps to determine preferences represented by a network map. Other tools such as SENSE are designed for
intent and large flow transfers over multi-domain networks. All of these approaches assume complete knowledge of the topology and work well in small networks. Extending bandwidth reservation tools such as OSCARS \cite{4374316} have been used to provide multi-layer network control directing traffic to flow over strategically planned routes. Using VLANs and intent-based and network markup language, application needs can be used to tailor paths. However, this approach requires precomputation and is often difficult to readjust if traffic patterns change. Here SR can provide good alternatives to table-based routing, where we can have more rules specified. In particular, SR solutions have been shown  to implement multipath telemetry solutions using P4, to implement data plane path reconfiguration \cite{mpolka}. 

Using ML for TE is a powerful way to provide decentralized decision-making, but requires engineering APIs. In this work we make progress towards this idea by exposing Hecate APIs for TE tools that can use these to make decisions. 

The use of ML and IA to improve TE on an SDN \cite{8870277} with MPLS and Segment Routing to create tunnels using the SR advantages, propose a hybrid paradigm with centralized ML and distributed IA on routers. The prediction of traffic was explored \cite{10339689} and implemented with Segment Routing; the protocol divides the network into segments, and each one has a Segment Identifier (SID), and this information is inserted into the packets. Shortest path routing techniques, like OSPF, are used to route the packet within the segment. The PolKA protocol can specify all the nodes in the path without increasing the header like MPLS does.


\section{Conclusion and Future work}
\label{sec:conclusion_n_future_work}



SDNet will transform computing networks by providing seamless control, monitoring, and automation of network applications. From a path-aware network perspective, we explore the PolKA architecture \cite{dominicini2020polka}, which introduces an innovative approach by leveraging the polynomial residue number system instead of traditional segment routing that relies on port switching. The PolKA routing algorithm, employs a highly efficient packet forwarding method based on the remainder of division, enabling stateless core nodes with a traffic control from the edge nodes. In this paper, we presented an integrated real-time framework that combines machine learning and optimization modules to enhance PolKA's traffic allocation across multiple network paths. This framework is deployable on both SDNet physical and emulated testbeds, ensuring efficient and seamless path selection.

In the future, we will be building upon this work and experimenting with more machine learning models such as neural networks, autoencoders and deep reinforcement learning techniques. The most immediate is to integrate the proposed framework for real-time source routing to evaluate path selection performance on a real testbed, such as a P4 lab or a continent-wide topology scenario \cite{wtestbeds2023}. For example, this could involve extending and deploying Hecate-PolKA framework into the Fabric testbed to assess the transfer of data-intensive scientific applications \cite{Performance-DIS-sbrc2024}. With the other promising direction of exploring advanced AI models, we will include deep neural networks and time series estimation models, to improve route selection and optimization in source routing with PolKA. Additionally, investigating energy efficiency e.g. by removing the table lookup from switches, which leads to reduced power consumption \cite{keyflow-energy} and resource-aware AI/ML models for the PolKA routing is crucial. 


\section*{Acknowledgments}

This work was supported by the U.S. DOE Office of Science, Office of Advanced Scientific Computing Research Early Career Grant ``Large Scale Deep Learning for Intelligent Networks'' award ERKJ435 hosted at Oak Ridge National Laboratory. This manuscript has been authored by UT-Battelle, LLC, under contract DE-AC05-00OR22725 with the US Department of Energy (DOE). The US government retains and the publisher, by accepting the article for publication, acknowledges that the US government retains a nonexclusive, paid-up, irrevocable, worldwide license to publish or reproduce the published form of this manuscript or allow others to do so, for US government purposes. DOE will provide public access to these results of federally sponsored research in accordance with the DOE Public Access Plan (https://www.energy.gov/downloads/doe-public-access-plan). Further, PolKA, is supported via Financial support from Brazilian agencies: CNPq, CAPES, FAPESP/MCTI/CGI.br, PORVIR-5G 20/05182-3, FAPES (94/2017, 281/2019, 515/2021, 284/2021, 06/2022, 1026/2022, 941/2022). CNPq fellows Dr. Martinello 306225/2020-4. SFI 13/RC/2077\_p2 and 17/CDA/4760.

\bibliographystyle{ieeetr}
\bibliography{./biblio.bib}

\end{document}